\documentstyle[PASJadd]{PASJ95}
\def\NH3{NH$_3$}

\begin{document}
\setcounter{page}{1}

\title{Observations of Ammonia in External Galaxies. 
II. Maffei 2}

\author{Shuro {\sc Takano} and Naomasa {\sc Nakai}\\  
{\it Nobeyama Radio Observatory (NRO)
\thanks{Nobeyama Radio Observatory is a branch of the National
Astronomical Observatory, an inter-university research institute 
operated by the Ministry of Education, Science, Sports, and Culture. 
This work was carried out under the common use observation program 
at NRO.},}\\
{\it Nobeyama, Minamimaki, Minamisaku, Nagano 384-1305}\\
{\it E-mail(ST): stakano@nro.nao.ac.jp}
\\[6pt]
Kentarou {\sc Kawaguchi}\\
{\it Department of Chemistry, Okayama University, 
Tsushima-naka 3-1-1,}\\ 
{\it Okayama 700-8530}\\
and\\
Toshiaki {\sc Takano}\\
{\it Graduate School of Science and Technology, Chiba University, 
Yayoi 1-33, Inage,}\\
{\it Chiba 263-8522}\\
}
\abst{The ammonia ($J$,$K$) = (1,1), (2,2), (3,3), and (4,4) 
transitions at 23.7 -- 24.1 GHz region were
searched for in a nearby galaxy Maffei 2
to study relation between molecular abundances 
and physical conditions in galaxies. 
The (1,1), (2,2), and (3,3) emission lines were clearly detected.
The rotational temperatures and ortho-to-para abundance ratios obtained
are about 30 K and about 2.6, respectively.
%The ratio of 2.6 corresponds to a temperature of distribution between 
%ortho- and para-\NH3 of about 13 K suggesting grain surface formation.
The abundance of \NH3 relative to H$_2$
in Maffei 2 was found to be 
the largest among galaxies where \NH3 is already detected, and
the abundance in Maffei 2 is more than an order of magnitude 
larger than the already reported upper limit in M82. 
Hence, we further confirmed the systematically 
peculiar molecular abundance in the aspect of formation 
mechanisms of molecules already reported
in M82.}

\kword{atomic and molecular processes --- 
galaxies: abundances --- galaxies: individual (Maffei 2) --- 
galaxies: starburst --- interstellar: molecules}

\maketitle
\thispagestyle{headings}

% section 1 ###########################################################
\section
{Introduction}

About 120 molecules have been detected in interstellar space and
circumstellar envelopes of late-type stars.
One fifth of them, 24 molecules, have already been detected in 
external galaxies (e.g. Henkel et al. 1991). 
%A few very abundant molecules such as CO have been used to study
%distributions and kinematics of molecular gas in galaxies.
%On the other hand, other less abundant
%molecules have been detected and used to study 
%physical conditions and chemical reactions in nearby galaxies
%with rich molecular gas in this $\sim$ 10 years.  
Further detections of molecules in galaxies, and subsequently, 
studies of relation between the molecular 
abundances and the physical conditions are 
important to understand chemical processes in galaxies.
Such knowledge is a 
basis to understand physics and evolution of galaxies 
by observations of molecular lines.
\par
In particular, molecular abundances of two well-known starburst 
galaxies, NGC 253 and M82, have been extensively studied (e.g. 
Henkel, Mauersberger 1992).
They have very rich and similar amount of molecular gas
(H$_2$ column density $\sim$ 10$^{23}$ cm$^{-2}$), and their 
distances from our Galaxy are nearly the same (about 3 Mpc). 
Consequently, they are good sources to compare the molecular abundances 
each other. 
As a result, clear difference in the molecular abundances has
been found.
Six molecules, SO, SiO, \NH3, HNCO, CH$_3$OH, and CH$_3$CN, 
have been clearly detected in NGC 253, 
but not detected with comparable sensitivity 
in M82 (Henkel et al. 1987; 
Mauersberger et al. 1991; 
Mauersberger, Henkel 1991; 
Nguyen-Q-Rieu et al. 1991; 
Takano et al. 1995; Takano et al. 2000). 
Actually the abundances of these six molecules are significantly 
higher in 
NGC 253 than those (upper limits) in M82. 
We noticed that these six molecules have common characteristics 
on their formation mechanisms; their formations are related to grain
and they are evaporated (or sputtered) from grain to gas phase, and/or
they are efficiently produced in gas-phase in active regions such as shock
regions and high-temperature regions (Takano et al. 1995, 2000).
Furthermore, at least half of above six molecules, \NH3, HNCO, and CH$_3$OH, 
is more abundant not only in NGC 253 but in nearby galaxies 
with rich molecular gas than in M82.
Based on these results, we concluded that molecular abundance in M82
is systematically peculiar in the aspect of formation mechanisms of 
molecules (Takano et al. 2000).
\par
\NH3 is already detected in NGC 253 and IC 342 in external galaxies
(Martin, Ho 1979; Takano et al. 2000).
Further detections of \NH3 in external galaxies will contribute
to the confirmation of the peculiarity in M82.
In addition, \NH3 is a good probe to obtain important physical
parameters such as temperature and ortho-to-para ratio, because
lines with different excitations and different spin statistical
species can be simultaneously observed
around 23 GHz region.
\par
In our preliminary survey of \NH3 in nearby galaxies, we detected
possible features in a galaxy Maffei 2 with the Effelsberg 100-m radiotelescope
in October 1998.
Maffei 2 is a nearby ($\sim$ 5 Mpc, Allen, Raimond 1972; Spinrad et al. 1973) 
barred galaxy [SBb(s) pec, Hurt et al. 1993a] obscured
by our Galactic disk ($b \sim -0^{\circ}\hspace{-4.5pt}.\hspace{.5pt}3$).
%and$l = 136^{\circ}$)
It has strong CO emissions (e.g. Weliachew et al. 1988), 
and detections of HNCO and CH$_3$OH among above six molecules
have already been reported (Nguyen-Q-Rieu et al. 1991; H\"uttemeister
et al. 1997).
We, therefore, searched for \NH3 deeply in Maffei 2, and
detections are reported in this letter.
%
% section 2 ##########################################################
\section
{Observations}

The observations were made with the NRO 45-m 
radiotelescope in 2000 June 
1 to 9. 
The adopted central position is
$\alpha$(1950) = $02^{\rm h}38^{\rm m}07^{\rm s}\hspace{-
5pt}.\hspace{2pt}98$ and
$\delta$(1950) = $59^{\circ}23'24.\hspace{-2pt}''8$.
This is a peak of K-band image (H. Spinrad, private communication). 
The systemic velocity (LSR) is --10 km s$^{-1}$.
Four inversion transitions of NH$_3$, (1,1), (2,2), (3,3), 
and (4,4)
at 23694.495, 23722.633, 23870.129, and 24139.417 MHz, respectively, 
were observed simultaneously using a 
dual polarization HEMT amplifier receiver.
Both circular polarizations were observed simultaneously.
The system noise temperature was 150 -- 
260 K (single side band) depending on elevation and weather. 
The main beam efficiency ($\eta_{mb}$) was 0.82, and the beam 
size was 70" (FWHM), which corresponds to about 1.7 kpc 
at the distance of 5 Mpc.
\par
The backend used was eight acousto-optical spectrometers. 
The band width is 40 MHz, and the resolution is 37 kHz for each 
spectrometer. 
The frequency resolution corresponds to 0.47 km s$^{-1}$ at 23.7 
GHz. 
The observations were made in a position switching mode. 
The off position was ($\Delta\alpha$, $\Delta\delta$) = 
($0.\hspace{-2pt}'5$, --12$'$),
which is free from foreground Galactic lines (Rickard et al. 1977).
The pointing was checked every $\sim$ 4 hours by observing an SiO 
maser line (\it J \rm = 1 -- 0, \it v \rm = 1) from a nearby 
evolved star S Per; 
average of each pointing deviation was 7". 
The calibration of the line intensity was made using the 
chopper-wheel method. 
%The obtained antenna temperature, $T^*_A$, was corrected for 
%atmospheric and 
%antenna ohmic losses, but not for the beam efficiency. 
%
%==================================================================
% section 3
\section
{Results and Analysis}

Figure 1 shows the obtained spectra.
The (1,1), (2,2), and (3,3) emission lines of \NH3 were clearly
detected, but the (4,4) line is only marginally detected at the 
$V_{LSR} \sim$ 15 km s$^{-1}$.  
%%%%%%%%%%%%%%%%%%%%%%%%%%%
%
%   Figure 1  
%
%%%%%%%%%%%%%%%%%%%%%%%%%%%
The lineshapes of the (1,1), (2,2), and (3,3) lines
are similar each other and show double peaks at the $V_{LSR}$ of about 
--90 and 10 km s$^{-1}$.
The intensities of the (1,1) and (3,3) lines are similar, but
the intensity of the (2,2) line is about half of them.
The line parameters are listed in table 1 for the total velocity range
and for each velocity component
($V_{LSR}$ = --150 -- --50 km s$^{-1}$ and --50 -- 100 km s$^{-1}$).
%%%%%%%%%%%%%%%%%%%%%%%%%%%
%
%   Table 1
%
%%%%%%%%%%%%%%%%%%%%%%%%%%%
\par
The analysis of the spectra was done with the similar method as
in the case of \NH3 in NGC 253 (Takano et al. 2000).
The rotational temperature from the (1,1) and (2,2) lines,
and the ortho-to-para ratio from the (1,1), (2,2), and (3,3) lines
are obtained by analyzing the line integrated intensity with the 
rotation diagram method (e.g. Turner 1991).
The beam filling factor was estimated to be 0.039, where the distribution 
of \NH3 was assumed to be the same as CO.  
The size of the CO distribution employed was 30" $\times$ 7" (FWHM)
from the observations of Ishiguro et al. (1989) and Hurt et al.
(1993b).
%A formula of convolution between two dimentional Gaussian and a telescope
%beam was used for the calculation (e.g. Burton et al. 1991).
The estimated value of the filling factor affects evaluation of 
column densities, but not for rotational temperatures and ortho-to-para
ratios.
The rotation diagram is shown in figure 2.
%%%%%%%%%%%%%%%
%
%   Figure 2
%
%%%%%%%%%%%%%%%
The obtained rotational temperatures are about 30 K for the total
velocity range and for the two velocity components.
The column densities are 1.5 $\times$ 10$^{15}$ cm$^{-2}$, 5.0
$\times$ 10$^{14}$ cm$^{-2}$, and 1.1 $\times$ 10$^{15}$ cm$^{-2}$
for the total velocity range, --150 -- --50 km s$^{-1}$, 
and --50 -- 100 km s$^{-1}$, respectively.
In addition the column density ratios of ortho- and para-\NH3  
([ortho]/[para], hereafter "ortho-to-para abundance 
ratio") are calculated to be 2.6$^{+1.1}_{-0.8}$, 
2.6$^{+2.8}_{-1.3}$, and 2.6$^{+1.4}_{-0.9}$ for above
velocity ranges, respectively. 
The errors correspond to $\sim$3$\sigma$.
These values are significantly larger than the value of
high temperature limit of 1.0 (see section 4.2.).
The obtained parameters are listed in table 2.
%%%%%%%%%%%%%%%%%%%%%%%%%%%
%
%   Table 2
%
%%%%%%%%%%%%%%%%%%%%%%%%%%%
%section 4  ###############################################################
\section{Discussion}
% subsection 4.1
\subsection{The Abundances of \NH3 in the Galaxies}
The abundance of \NH3 in Maffei 2 relative to H$_2$ ([\NH3]/[H$_2$]) 
is calculated from their column densities.
As the column density of H$_2$, we employed 1.4 $\times$ 10$^{22}$ 
cm$^{-2}$ (H\"uttemeister et al. 1995).
The obtained abundance is 1 $\times$ 10$^{-7}$, which is higher
than the corresponding values of 2.7 $\times$ 10$^{-8}$ in NGC 253
and 1.9 $\times$ 10$^{-8}$ in IC 342 (Takano et al. 2000).
Maffei 2 has the highest abundance of \NH3 among galaxies where \NH3
is detected.
\par
Abundances of \NH3 in our 
Galactic sources strongly depend on regions. 
The abundances are between 1$\times$10$^{-8}$ and 10$^{-4}$ as follows.
In our Galactic center Sgr B2 the abundances are reported to be 
(1 -- 10)$\times$10$^{-8}$ (Irvine et al. 1987) and 8$\times$10$^{-
8}$ -- 10$^{-4}$ (H\"uttemeister et al. 1993).
In the cyanopolyyne peak of TMC-1, which is a quiescent young dark cloud
where ammonia formation is not yet prominent, the abundance is 
2$\times$10$^{-8}$ (Irvine et al. 1987).
In star forming regions abundances are typically more than  
10$^{-7}$.
For example the abundance in the Orion ridge
is 2$\times$10$^{-7}$  (Irvine et al. 1987).
The abundance in Maffei 2 
is similar to this value. 
\par
The abundance in Maffei 2 is much higher than the upper limit
of 1.4 $\times$ 10$^{-9}$ in M82 (at T$_{rot}$ = 30 K, Takano et al. 2000), 
though star formation activity and amount of molecular gas
of Maffei 2 are significantly lower than those in M82;
peculiarity of the molecular abundance in M82 is further supported
by the present result.
%subsection 4.2 #######################################################
\subsection{Ortho-to-para Abundance Ratios of \NH3}
There are ortho-\NH3 ($K$ = 3n, n is integer) 
and para-\NH3 ($K$ $\not=$ 3n) due to the three equivalent  
hydrogen nuclei with spin 1/2. 
No dipole and collisional transitions are allowed between the ortho 
and para states.
The spin statistical weight is 4 : 2 for ortho : para (e.g. Townes,
Schawlow 1975). 
On the other hand, the number of the para levels is almost two times larger 
than that 
of the 
ortho levels, and consequently the ortho-to-para abundance ratio 
approaches to 1.0, if \NH3 is produced and equilibrated under 
high-temperature 
conditions ($\gtsim$ 40 K). 
On the other hand, the ratio becomes quite large if \NH3 is 
produced and equilibrated under low-temperature conditions (e.g. 
more than 10 
at 5 K), because the lowest level, (0,0), belongs to ortho. 
The dependence of ortho-to-para abundance ratios of \NH3 on
temperatures, which determine the distribution between ortho and para, 
was calculated and shown in figure 3 of Takano et al. (2000).
\par
The ortho-to-para abundance ratios obtained in Maffei 2 are
about 2.6 for all velocity components.
%This ratio is higher than 1.0, which is a value at high-temperature
%limit.
The ratio at the total velocity component is 2.6$^{+1.1}_{-0.8}$,
which corresponds to a temperature of distribution between
ortho- and para-\NH3 of 13$^{+5}_{-3}$ K.
If this temperature can be  interpreted as a formation
temperature of \NH3, it was produced on low-temperature grain
and subsequently evaporated to gas-phase.
\par
Studies of \NH3 abundance in the aspect of an ortho-to-para abundance
ratio have been limited.
In our Galaxy, Umemoto et al. (1999) recently obtained ratios of
1.3 -- 1.7, which are significantly higher than the value of
high-temperature limit of 1.0, in the outflow lobe of a star 
forming region L1157.
In external galaxies, ratios have been obtained only in NGC 253
(Takano et al. 2000). 
The values are quite different depending on the velocity components;
$\sim$ 1 for the relatively red-shifted component and more than $\sim$
14 for the relatively blue-shifted component.
The ratio obtained in Maffei 2 is not so large as obtained in the relatively
red-shifted component in NGC 253. 
%
%subsection 4.3 ###############################################
\subsection{Double Peak Spectra of \NH3}
As shown in figure 1, the lineshapes of the \NH3 lines in Maffei 2 
show double peak.
Such double peak has already been reported at the CO 
$J$ = 1 -- 0 (Sargent et al. 1985; Weliachew et al. 1988), 
$J$ = 2 -- 1 (Sargent et al. 1985), and
$J$ = 3 -- 2 (Hurt et al. 1993b; Mauersberger et al. 1999) transitions.
In addition, the H$_2$CO 2$_{11}$ -- 1$_{10}$ and 2$_{12}$ -- 1$_{11}$
transitions (H\"uttemeister et al. 1997), and
HNC $J$ = 1 -- 0 (H\"uttemeister et al. 1995) transition also
show the double peak. 
Ishiguro et al. (1989) found a molecular ring 
with a diameter of about 500 pc $\times$ 240 pc in the bar.
Gas is deficient inside of the ring.
These double peaks can be interpreted to originate from this
ring structure.
%These molecules including \NH3 are, therefore, abundant in the
%central region of Maffei 2.
For CO and \NH3, the velocity component at $\sim$ 10 km s$^{-1}$ is clearly
stronger than that at $\sim$ --90 km s$^{-1}$.
This may be related to the small scale asymmetries in the molecular bar
(Ishiguro et al. 1989) and to the large scale asymmetry of the arms found
by the infrared observation (Hurt et al. 1993a).
%==============================================================
\par
\vspace{1pc}\par
We thank all members of the 45-m telescope in Nobeyama 
for support of our observations. 
We are grateful to H. Spinrad for informing us the coordinate 
before publication.
We thank P. Schilke for his support in the preliminary survey 
of \NH3 with the Effelsberg 100-m telescope.

\section*{References}
\small

% checked
\re
Allen R.J., Raimond E.\ 1972,  A\&A 19, 317

%\re
%Adams, N.G., Smith, D., Millar, T.J.\ 1984,  MNRAS 211, 857

%\re
%Bachiller, R., Mart{\'\i}n-Pintado, J., Fuente, A.\ 1993, ApJ 417, 
%L45

%\re
%Becklin, E.E., Gatley, I., Matthews, K., Neugebauer, G., Sellgren, K.,
%Werner, M.W., Wynn-Williams, C.G.\ 1980, ApJ 236, 441

%\re
%Bettens, R.P.A., Collins, M.A.\ 1998, J.Chem.Phys. 109, 9728

%\re
%Brown, P.D., Charnley, S.B.\ 1990, MNRAS 244, 432

%\re
%Canzian, B., Mundy, L.G.,  Scoville,  N.Z.\ 1988, ApJ 333, 157

%\re
%Cesaroni, R., Walmsley, C.M., Churchwell, E.\ 1992, A\&A 256, 618

%\re
%Cheung, A.C., Rank, D.M., Townes, C.H., Knowles, S.H., 
%Sullivan III, W.T.\ 1969, ApJ 157, L13

%\re
%Coles, D.K., Good, W.E., Bragg, J.K., Sharbaugh, A.H.\ 1951, 
%Phys.\ Rev. 82, 877

%\re
%Cottrell, G.A.\ 1977, MNRAS 178, 577
%
%\re
%Davidge, T.J., Le F\`evre, O., Clark, C.C.\ 1991, ApJ 370, 559
%
%\re
%Devereux, N.A., Young, J.S.\ 1990, ApJ 359, 42

%\re
%d'Hendecourt, L.B., Allamandola, L.J., Greenberg, J.M.\ 1985, A\&A 
%152, 130

%\re
%Galloway, E.T., Herbst, E.\ 1989, A\&A 211, 413

%\re
%Gaume, R.A., Johnston, K.J., Nguyen, H.A., Wilson, T.L., Dickel, 
%H.R., 
%Goss, W.M., Wright, M.C.H.\ 1991, ApJ 376, 608

\re
Henkel C., Baan W.A., Mauersberger R.\ 1991, A\&AR 3, 47

\re
Henkel C., Jacq T., Mauersberger R., Menten K.M., Steppe H.\ 
1987, A\&A 188, L1

\re
Henkel C., Mauersberger R.\ 1992,  Astrochemistry of Cosmic 
Phenomena, ed P.D. Singh (Kluwer Academic Publishers, Dordrecht) 
p111

%\re
%Henkel C., Whiteoak J.B., Mauersberger R.\ 1994, A\&A 284, 17

%\re
%Herbst, E., DeFrees, D.J., McLean, A.D.\ 1987, ApJ 321, 898

%\re
%Herbst, E., Klemperer, W.\ 1973, ApJ 185, 505

%\re
%Hiraoka, K., Yamashita, A, Yachi, Y., Aruga, K., Sato,T., Muto, H.\ 
%1995, ApJ 443, 363

%\re
%Ho, P.T.P., Martin, R.N.\ 1983, ApJ 272, 484

%\re
%Ho, P.T.P., Martin, R.N., Turner, J.L., Jackson, J.M.\ 1990, ApJ 
%355, L19

%\re
%Ho, P.T.P., Townes, C.H.\ 1983, ARA\&A 21, 239 

%\re
%Hofner, P., Kurtz, S., Churchwell, E., Walmsley, C.M., Cesaroni, 
%R.\ 
%1994, ApJ 429, L85

% checked
\re
Hurt R.L., Merrill K.M., Gatley I., Turner J.L.\
1993a, AJ 105, 121

% checked
\re
Hurt R.L., Turner J.L., Ho P.T.P., Martin R.N.\
1993b, ApJ 404, 602

\re
H\"uttemeister S., Henkel C., Mauersberger R., Brouillet N.,
Wiklind T., Millar T.J.\
1995, A\&A 295, 571

\re
H\"uttemeister S., Mauersberger R., Henkel C.\
1997, A\&A 326, 59

\re
H\"uttemeister S., Wilson T.L., Henkel C., Mauersberger R.\
1993, A\&A 276, 445

%\re
%Ichikawa, T., Yanagisawa, K., Itoh, N., Tarusawa, K., van Driel, 
%W., and Ueno, M.\ 1995, AJ 109, 2038 

%\re
%IRAS Point Source Catalog, 1988, The Joint IRAS Science Working 
%Group,
%NASA (Washington, DC)

%\re
%IRAS Science Team, 1986, A\&AS 65, 607

\re
Irvine W.M., Goldsmith P.F., Hjalmarson \AA.\  1987,   
Interstellar Processes, ed D.J. Hollenbach, H.A. Thronson, Jr.\
 (D. Reidel Publishing Company, Dordrecht) p561

\re
Ishiguro M., Kawabe R., Morita K.-I., Okumura S.K., Chikada 
Y., Kasuga T., Kanzawa T., Iwashita H. et al.\ 1989, ApJ 344, 763

%\re
%Ishizuki, S., Kawabe, R., Ishiguro, M., Okumura, S.K., Morita, K.-
%I., et al.\ 
%1990, Nature 344, 224

%\re
%Israel, F.P.\  1992, A\&A 265, 487

%\re
%Jackson, J.M., Paglione, T.A.D., Carlstrom, J.E., Nguyen-Q-Rieu 
%1995, ApJ 438, 695

%\re
%Kalas, P., Wynn-Williams, C.G.\ 1994, ApJ 434, 546

%\re
%Kraemer, K.E., Jackson, J.M.\ 1995, ApJ 439, L9

%\re
%Kroto, H.W.\ 1992, Molecular Rotation Spectra (Dover, New York), p58

%\re
%Kr\"ugel, E., Chini, R., Klein, U., Lemke, R., Wielebinski, R., 
%Zylka, R.\
%1990, A\&A 240, 232

%\re
%Le Bourlot, J.\ 1991, A\&A 242, 235

%\re
%Lilley, A.E., Palmer, P.\ 1968, ApJS 16, 143 

%\re
%Lo, K.Y., Cheung, K.W., Masson, C.R., Phillips, T.G., Scott, S.L., 
%Woody, D.P.\ 1987, ApJ 312, 574

%\re
%Loinard,L., Allen, R.J.\ 1998, ApJ 499, 227

%\re
%Mangum, J.G., Wootten, A.\ 1994, ApJ 428, L33

%\re
%Marquette, J.B., Rebrion, C., Rowe, B.R.\ 1988, J.\ Chem.\ Phys.\ 
%89, 2041

\re
Martin R.N., Ho P.T.P.\ 1979, A\&A 74, L7

%\re
%Martin, R.N., Ho, P.T.P.\ 1986, ApJ 308, L7

%\re
%Martin, R.N., Ho, P.T.P., Ruf, K.\ 1982, Nature 296, 632

\re
Mauersberger R., Henkel C.\ 1991, A\&A 245, 457

%checked
\re
Mauersberger R., Henkel C., Walmsley C.M., Sage L.J., Wiklind
T.\ 1991, A\&A 247, 307

%checked
\re
Mauersberger R., Henkel C., Walsh W., Schulz A.\ 1999, A\&A 341, 256

%\re
%Mauersberger R., Henkel C., Sage, L.J.\ 1990, A\&A 236, 63

%\re
%Mauersberger R., Henkel C., Wielebinski, R., Wiklind, T., Reuter, 
%H.-P.\ 1996, A\&A 305, 421

%\re
%Mauersberger R., Wilson, T.L., Henkel C.\ 1986, A\&A 160, L13

%\re
%McCall, M.L.\ 1989, AJ 97, 1341

%\re
%Menzel, D.H.\ 1969, ApJS 18, 221

%\re
%Nakai N., Hayashi, M., Handa, T., Sofue, Y., Hasegawa, T., 
%Sasaki, M.\ 1987, PASJ 39, 685

\re
Nguyen-Q-Rieu, Henkel C., Jackson J.M., Mauersberger R.\ 1991, 
A\&A 241, L33

%\re
%Pi$\tilde{\rm n}$a, R.K., Jones, B., Puetter, R.C.\ 1992, 
%ApJ 401, L75

%\re
%Peng, R., Zhou, S., Whiteoak, J.B., Lo, K.Y., Sutton, E.C.\ 1996, 
%ApJ 470, 821

%\re
%Pipher, J.L., Moneti, A., Forrest, W.J., Woodward, C.E., Shure, 
%M.A.\ 1987, Infrared Astronomy with Arrays, ed C.G.\ Wynn-Williams, 
%E.E.\ Becklin (University of Hawaii, Honolulu) p326

%\re
%Puxley, P.J., Brand, P.W.J.L., Moore, T.J.T., Mountain, C.M., 
%Nakai N., Yamashita, T.\ 1989, ApJ 345, 163

%\re
%Puxley, P.J., Mountain, C.M., Brand, P.W.J.L., Moore, T.J.T., 
%Nakai N.\ 1997, ApJ 485, 143

\re
Rickard L.J., Turner B.E., Palmer P.\ 1977, ApJ 218, L51

%\re
%Rieke, G.H., Lebofsky, M.J., Thompson, R.I., Low, F.J., and 
%Tokunaga, A.J.\ 1980, ApJ 238, 24

%\re
%Rieke, G.H., Lebofsky, M.J., Walker, C.E.\ 1988, ApJ 325, 679

%\re
%Rowe, B.R., Smith, I.W.M.\ 1992,  J.\ Chem.\ Phys.\  97, %8798

\re
Sargent A.I., Sutton E.C., Masson C.R., Lo K.Y., Phillips
T.G.\ 1985, ApJ 289, 150

%\re
%Scott, G.B.I., Freeman, C.G., McEwan, M.J.\ 1997, MNRAS 290, 636

%\re
%Seaquist, E.R., Kerton, C.R., Bell, M.B.\ 1994, ApJ 429, 612

%checked
\re
Spinrad H., Bahcall J., Becklin E.E., Gunn J.E., Kristian J.,
Neugebauer G., Sargent W.L.W., Smith H.\ 1973, ApJ 180, 351

%checked
%\re
%Spinrad, H., Sargent, W.L.W., Oke, J.B., Neugebauer, G., Landau, R.,
%King, I.R., Gunn, J.E., Garmire, G., Dieter, N.H.\ 1971, ApJL 163, 25

%\re
%Suzuki, H., Yamamoto, S., Ohishi, M., Kaifu, N., Ishikawa, S., 
%Hirahara, Y., 
%Takano S.\ 1992, ApJ 392, 551

%\re
%Tafalla, M., Bachiller, R.\ 1995, ApJ 443, L37

\re
Takano S., Nakai N., Kawaguchi K.\ 1995, PASJ 47, 801

\re
Takano S., Nakai N., Kawaguchi K.\ 2000, submitted to PASJ 

%\re
%Tammann, G.A., Sandage, A.\ 1968, ApJ 151, 825

\re
Townes C.H., Schawlow A.L.\ 1975, Microwave Spectroscopy (Dover, 
New York), p 72

\re
Turner B.E.\ 1991, ApJS 76, 617

%\re
%Turner, J.L., Ho, P.T.P.\ 1983, ApJ 268, L79

%\re
%Turner, J.L., Ho, P.T.P.\ 1985, ApJ 299, L77

%\re
%Turner, J.L., Hurt, R.L.\ 1992, ApJ 384, 72

%\re
%Turner, J.L., Hurt, R.L., Hudson, D.Y.\ 1993, ApJ 413, L19

%\re
%Ulvestad, J.S., Antonucci, R.R.J.\ 1997, ApJ 488, 621

%\re
%Umemoto, T., Iwata, T., Fukui, Y., Mikami, H., Yamamoto, S., 
%Kameya, O., Hirano, N.\ 1992, ApJ 392, L83

\re
Umemoto T., Mikami H., Yamamoto S., Hirano N.\ 1999, ApJ 525, L105

%\re
%Vikor, L., Al-Khalili, A., Danared, H., Djuric, N., Dunn, G.H., Larsson, M.,
%Le Padellec, A., Ros\'en. S., af Ugglas, M.\ 1999, A\&A 344, 1027

%\re
%Walmsley, C.M., Hermsen, W., Henkel C., Mauersberger R., Wilson, 
%T.L.\ 1987, A\&A 172, 311

\re
Weliachew L., Casoli F., Combes F.\ 1988, A\&A 199, 29

%\re
%Yee, J.H., Lepp, S., Dalgarno, A.\ 1987, MNRAS 227, 461

%\re
%Yun, M.S.\ 1999, in Galaxy Interactions at Low and High Redshift (IAU symposium 186), 
%ed J.E. Barnes, D.B. Sanders (Kluwer Academic Publishers, Dordrecht) 
%p81

%\re
%Zhang, Q., Ho, P.T.P.\ 1995, ApJ 450, L63

\label{last}

% Table 1
\clearpage
\begin{table}
\small
\begin{center}
Table~1.\hspace{4pt}Observed line parameters of the ammonia (1,1), 
(2,2), 
(3,3), and (4,4) transitions.\\
\end{center}
\vspace{6pt}
\begin{tabular*}{\textwidth}{@{\hspace{\tabcolsep}
\extracolsep{\fill}}p{5pc}cccccc}
\hline\hline\\[-6pt]
 Transition & Velocity range & $\smallint T_{mb}dv$ 
$^{a}$ & 
$T_{mb}$ $^{b}$ & $V_{LSR}$ $^{b}$ & $\Delta$v (FWHP) $^{b}$ & Rms noise 
($T_{mb}$) $^{c}$ \\
         & (km s$^{-1}$) & (K km s$^{-1}$) & (mK) & (km s$^{-1}$) & (km s$^{-1}$) 
         & (mK) \\
[4pt]\hline\\[-6pt]
(1,1) \dotfill & --150 -- 100 (total) & 0.89$\pm$0.06 & 7.4 & $\cdots$ & 
$\cdots$ & 1.1 \\
      & --150 -- --50         & 0.29$\pm$0.04 & 4.5 & --83 & 76 &     \\
      &  --50 -- 100          & 0.59$\pm$0.04 & 7.4 & 13   & 65 &     \\
(2,2) \dotfill & --150 -- 100 (total) & 0.50$\pm$0.04 & 4.9 & $\cdots$ & 
$\cdots$ & 0.9 \\
      & --150 -- --50        & 0.16$\pm$0.03 & 2.2 & --100 & 81 &     \\
      &  --50 -- 100         & 0.34$\pm$0.03 & 4.9 & 5     & 62 &     \\
(3,3) \dotfill & --150 -- 100 (total) & 0.95$\pm$0.04 & 6.5 & $\cdots$ & 
$\cdots$ & 0.9 \\
      & --150 -- --50        & 0.29$\pm$0.03 & 4.7 & --86 & 56  &     \\
      &  --50 -- 100         & 0.66$\pm$0.03 & 6.5 & 8    & 98  &     \\
(4,4) \dotfill &  --50 -- 100     & 0.06$\pm$0.05$^d$ & 3.3 &  15 & 
$\cdots$ & 1.2 \\
[4pt]
\hline
\end{tabular*}

\vspace{6pt}

\noindent
$^a$ $T_{mb} \equiv$ $T_A^*$/{$\eta_{mb}$} \ \ \ ($\eta_{mb}$ = 
0.82).
The error corresponds to 1$\sigma$.
\par\noindent
$^b$ Obtained by Gaussian fit except for the (4,4) line.
\par\noindent
$^c$ Velocity resolution is 10 km s$^{-1}$.
\par\noindent
$^d$  Only marginally detected.
\par\noindent

\end{table}

% Table 2
%%\clearpage
\begin{table}
\small
\begin{center}
Table~2.\hspace{4pt}Ammonia rotational temperatures,
column densities, ortho-to-para 
abundance ratios, and abundances.\\
\end{center}
\vspace{6pt}
\begin{tabular*}{\textwidth}{@{\hspace{\tabcolsep}
\extracolsep{\fill}}p{6pc}p{2pc}ccccc}
\hline\hline\\[-6pt]
  Velocity &T$_{rot}$ $^a$ &   & Column Density $^b$ & & Ortho-to-para $^c$
  & Abundance $^d$  \\
  Component &          & Ortho & Para & Total & Abundance Ratio &               \\
 (km s$^{-1}$) & (K) & (cm$^{-2}$) & (cm$^{-2}$) & (cm$^{-2}$) &    \\
[4pt]\hline\\[-6pt]
--150 -- 100 (total)\dotfill & 30$^{+3}_{-2}$   & (1.1$\pm$0.3) $\times$ 10$^{15}$  &
   (4.3$\pm$0.4) $\times$ 10$^{14}$  & (1.5$\pm$0.3) $\times$ 10$^{15}$
    & 2.6$^{+1.1}_{-0.8}$  & 1 $\times$ 10$^{-7}$ \\
--150 -- --50 \dotfill & 30$^{+5}_{-4}$  & (3.6$\pm$2.0) $\times$ 10$^{14}$  & 
   (1.4$\pm$0.3) $\times$ 10$^{14}$  & (5.0$\pm$2.0) $\times$ 10$^{14}$ 
    & 2.6$^{+2.8}_{-1.3}$ & $\cdots$ \\
--50 -- 100 \dotfill & 31$\pm$3    & (7.6$\pm$2.3) $\times$ 10$^{14}$   & 
   (2.9$\pm$0.3) $\times$ 10$^{14}$    & (1.1$\pm$0.2) $\times$ 10$^{15}$   
    & 2.6$^{+1.4}_{-0.9}$ & $\cdots$   \\
[4pt]
\hline
\end{tabular*}

\vspace{6pt}

\noindent
$^{a}$ Obtained from the populations between the (1,1) and (2,2)
levels. 
The error corresponds to $\sim$1$\sigma$.
\par\noindent
$^{b}$
The error corresponds to $\sim$1$\sigma$.
\par\noindent
$^{c}$
The error corresponds to $\sim$3$\sigma$.
\par\noindent
$^{d}$ The column density of H$_2$ (1.4 $\times$ 10$^{22}$ cm$^{-2}$)
is taken from  H\"uttemeister et al. (1995). 
\par\noindent

\end{table}

% Figure captions
%\clearpage
%\centerline{Figure Captions}
\bigskip
\begin{Fv}{1}{18pc}%
{Spectra in the frequency regions of the NH$_3$ 
(1,1), (2,2), 
(3,3), and (4,4) transitions from bottom to top. 
The LSR velocity shown is for each transition. 
The velocity resolution is 10 km s$^{-1}$. 
%The dashed line indicates the systemic velocity of Maffei 2 
%(--10 km s$^{-1}$).
The lines show double peak structure.
The (4,4) transition is only marginally detected.
}
\end{Fv}
\begin{Fv}{2}{18pc}%
{Rotation diagrams of the \NH3 lines.  
The points $\bullet$, $\circ$, and $\diamond$ indicate 
the data points of 
integrated intensity which covers the total velocity range 
(--150 -- 100 km s$^{-1}$), --150 -- --50 km s$^{-1}$, and 
--50 -- 100 km s$^{-1}$, respectively.
Lines connecting the (1,1) and (2,2) data points are shown.
} 
\end{Fv}
\end{document}